**Winding number correlation for a Brownian loop in a plane**

J.H. Hannay,
H.H.Wills Physics Laboratory, University of Bristol, Tyndall Ave, Bristol BS8 1TL, UK.

*Abstract*
A Brownian loop is a random walk circuit of infinitely many, suitably infinitesimal, steps. In a plane such a loop may or may not enclose a marked point, the origin, say. If it does so it may wind arbitrarily many times, positive or negative, around that point. Indeed, from the (long known) probability distribution, the mean square winding number is infinite, so all statistical moments - averages of powers of the winding number - are infinity (even powers) or zero (odd powers, by symmetry). If an additional marked point is introduced at some distance from the origin, there are now two winding numbers, which are correlated. That correlation, the average of the product of the two winding numbers, is finite, and is calculated here. The result takes the form of a single well-convergent integral that depends on a single parameter – the suitably scaled separation of the marked points. The integrals of the correlation weighted by powers of the separation are simple factorial expressions. Explicit limits of the correlation for small and large separation of the marked points are found.

*Introduction*
A mathematical 'Brownian motion' is an idealized random walk of infinitely frequent steps with infinitesimal step length such that the product *step length$^2$ × step frequency=finite constant*. The value of the constant can be chosen for convenience (choosing suitable units for time). Classic references for mathematical aspects of Brownian motion are [Ito and McKean 1965] and [Pitman and Yor 1986], and a more recent review is [Collet and Tourigny 2015]. In two dimensions a natural convention for the constant, used here, is 2. Then the mean square end-to-end displacement $(\Delta x, \Delta y)$ of a Brownian motion of any duration $\Delta t$ obeys $<\Delta x^2>=<\Delta y^2>=\Delta t$. A Brownian *loop* is a closed such random walk; its final point coincides with its initial point. Because of the infinity of steps, such a loop, suitably located, is capable of winding arbitrarily many times $n$ around some chosen marked point, say the origin, with $n$ any positive (anticlockwise), zero, or negative, integer. The study and central results of path winding really go back to the work in optics of Sommerfeld [Sommerfeld 1896, 1954][Hannay and Thain 2003]. A separate winding approach to the Aharonov-Bohm effect was taken by [Berry 1980]. If topology is being emphasized, the marked point is sometimes called a 'puncture' point – a point-like hole in the plane (as in [Giraud, Thain and Hannay 2004]).

The probability that a Brownian loop of finite duration $t$ located randomly in an entire plane encloses the origin is obviously infinitesimal – its winding number is almost certainly zero. The *conditional* probability for its winding number to be $n$, subject to this *not* being zero (i.e. not topologically detachable) is known to be $(\pi/n)^2/12$ independent of $t$ [Edwards 1967][Kleinert 2006][Giraud, Thain and

Hannay 2004].  The calculation runs, in outline, as follows.  Choose an arbitrary initial point for Brownian motion of duration $t$.  The ordinary Gaussian probability density for the location of the endpoint is given by (1).  For closed Brownian motion - a Brownian loop - the endpoint is back at the initial point and the probability density has the value $1/2\pi t$ there.  Split this, by one of a few possible equivalent procedures, into sub-probability densities according to the number $n$ of anticlockwise windings (positive, negative or zero) of the loop around the origin.  Finally, discarding the zero-winding class, integrate over all positions of the initial/final point and normalize this set of numbers to give the winding probabilities.   Often the moments of a set of probabilities– the average of the positive integer powers of $n$ are of interest.  However for Brownian loop windings around a single marked point they are not; by the clockwise/anticlockwise symmetry the average of any odd power of $n$ is zero, and, from the distribution stated, all even moments are infinite.  Because of the infinitely fine structure in Brownian paths, there is no finite-valued moment.

If a second marked point is introduced at a distance $L$ from the origin there is a second anticlockwise winding number $-m$ (the minus sign is a convention for later convenience, see Fig 1).  The 'winding number correlation' $\langle(-m)n\rangle$ is neither zero nor infinity, and it is to be calculated here.  By the scaling invariance of Brownian paths, it is a function of the single variable $L/\sqrt{t}$.  Higher winding moments may well not have finite values.  Those of the form $\langle(-m)^{even}n^{odd}\rangle$ and $\langle(-m)^{odd}n^{even}\rangle$ are zero by symmetry; ones of the form $\langle(-m)^{even}n^{even}\rangle$ are infinite if, as seems likely, the conditional expectation $\langle m^2|n\rangle$ is infinite for any given $n \neq 0$; finally higher moments of the form $\langle(-m)^{odd}n^{odd}\rangle$ are also infinite if $\langle m^3|n\rangle$ is infinite for any given $n \neq 0$ which seems plausible because of the infinitely fine structure in Brownian paths.

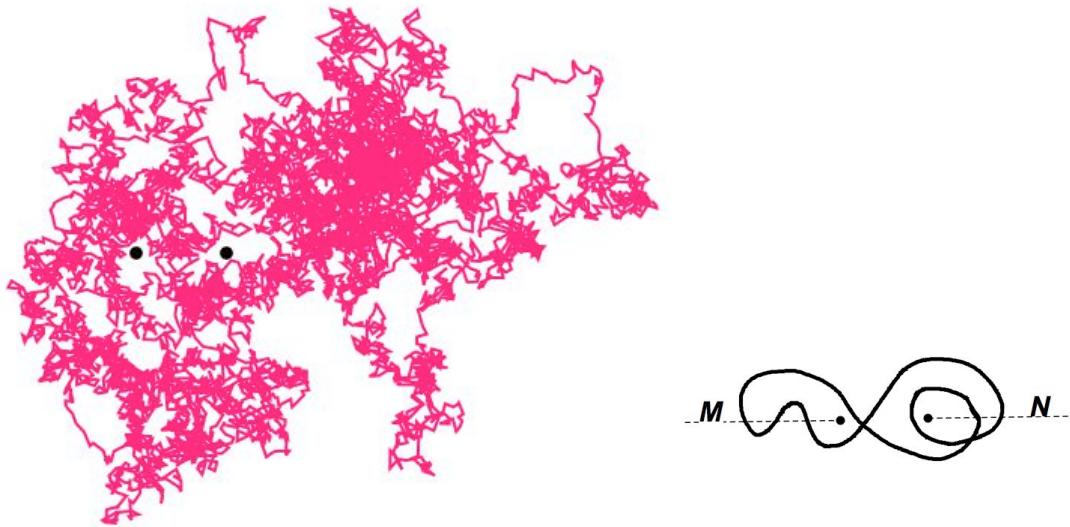

*Fig* 1.  (Left) A Brownian loop in the plane with two marked points.  The picture is obviously not well suited to visual counting of windings (Brownian motion is a fractal of dimension 2).  (Right) A schematic loop instead, to explain counting.  If one hypothetically moves around the loop so that the right hand dashed line *N* is crossed *n*=2 times in the upward direction, then the left hand dashed line *M* is crossed  *m*=1−1+1=1 times in the upward direction.  The winding numbers (anticlockwise) around left and right marked points would then be −*m*=−1 and

$n=2$ respectively. The product *mn* does not depend on the choice of which way around the loop to move.

*Calculation*

The method to be used is based on the ordinary Gaussian probability distribution for the end-to-end displacement ($\Delta x, \Delta y$) of a Brownian motion (not a loop) of duration $\Delta t$. With $<\Delta x^2>=<\Delta y^2>=\Delta t$ this is

$$P(\Delta x, \Delta y, \Delta t) \equiv \frac{1}{2\pi\Delta t} Exp\left[-\frac{\Delta x^2 + \Delta y^2}{2\Delta t}\right] \tag{1}$$

This obeys the diffusion equation with diffusion coefficient $D=\frac{1}{2}$ (that is, the probability flux density is half the probability density gradient).

$$\frac{\partial P}{\partial \Delta t} = \tfrac{1}{2}\frac{\partial^2 P}{\partial \Delta x^2} + \tfrac{1}{2}\frac{\partial^2 P}{\partial \Delta y^2} \tag{2}$$

Brownian motions of duration *t* from an arbitrary fixed initial point (*X,Y*) to a final point (*x,y*) that can be adjusted, are to be classified according to where they cross a hypothetical infinite straight line passing through the two marked points, possibly multiple times. Let the marked points lie at ±*L*/2 on the *x* axis (Fig 1), this is neater than having one at the origin. Crossings by the Brownian path through the segment of straight line between the marked points are immaterial, but two numbers, *m* and *n* count the net number of signed crossings through the two 'wing', half-line portions of the straight line, say *M* and *N* (fig 1). A crossing towards positive *y* is counted as a positive crossing and the reverse is negative. If, as later, the final point (*x,y*) is taken as coincident with the initial one (*X,Y*), the numbers (−*m*) and *n* obviously count the anticlockwise winding numbers of the Brownian loop around the two marked points. In principle any other choice for the directions of the half-lines *M* and *N* is equally valid, but the present choice is simplest operationally.

Define $P_{mn}(X,Y,0 \rightarrow x,y,t)$ as the sub-probability density for arriving at (*x,y*) from the fixed initial point (*X,Y*) after duration *t* via Brownian paths having the crossing numbers *m* and *n*. There is no simple formula for $P_{mn}$ (see the concluding remarks), but the winding number correlation to be calculated here does not require a formula for it. A key point is that $P_{mn}(X,Y,0 \rightarrow x,y,t)$, considered throughout as a function of the variable final position *x,y,t*, obeys the diffusion equation. Only on the half-line portions *M* and *N* does this not apply, there being a discontinuous jump of $P_{mn}(X,Y,0 \rightarrow x,y,t)$ across *M* and *N* because of the definitions of *m* and *n*. In contrast, the total of sub-probability densities is the free space probability density which has no jump:

$$\sum_{m,n} P_{mn}(X,Y,0 \rightarrow x,y,t) = P(x-X, y-Y, t) = \frac{1}{2\pi t} Exp\left[-\frac{(x-X)^2 + (y-Y)^2}{2t}\right]. \tag{3}$$

Quantities (real scalar fields, or more descriptively 'generating functions') of interest are:

$$Q_M(X,Y,0 \to x,y,t) \equiv \sum_{m,n} m\, P_{mn}(X,Y,0 \to x,y,t) \tag{4}$$

$$Q_N(X,Y,0 \to x,y,t) \equiv \sum_{m,n} n\, P_{mn}(X,Y,0 \to x,y,t) \tag{5}$$

$$Q_{MN}(X,Y,0 \to x,y,t) \equiv \sum_{m,n} m\, n\, P_{mn}(X,Y,0 \to x,y,t) \tag{6}$$

The first two can (and will) each be expressed as an integral in terms of $P$. The last will be expressed in terms of the first two and $P$. It is from the last that the desired winding correlation <–$mn$> is obtained by requiring coincidence between initial and final positions and integrating. The result then needs dividing by the measure of all allowed loop paths.

$$\langle mn \rangle = \frac{\iint Q_{MN}(x,y,0 \to x,y,t)\, dx\, dy}{\sum_{m,n} \iint (1-\delta_{n0}) P_{mn}(x,y,0 \to x,y,t)\, dx\, dy} \tag{7}$$

$$= 12 \iint Q_{MN}(x,y,0 \to x,y,t)\, dx\, dy$$

The denominator here supplies the measure, expressing the condition of topological non-detachment from the origin, by which the ensemble is defined (one might aspire to other definitions – see the concluding remarks). The sum over $m$ is trivial and the 'normalization' factor 12 then follows from the known single marked point analysis [Giraud, Thain and Hannay 2004]. (It should perhaps be noted that the factor $1-\delta_{n0}$ could equivalently be replaced by the more symmetrical $1-\tfrac{1}{2}\delta_{n0} - \tfrac{1}{2}\delta_{m0}$). The final result for <$mn$> will be reduced to a single integral expression (17).

Crucial to the argument will be the fact that, like $P_{mn}$, all the three quantities $Q$ as functions of $(x,y,t)$ obey the diffusion equation (2). They only differ through the boundary conditions. The boundary surface in 2+1D spacetime is shown in fig 2. All three quantities are zero at spatial infinity (shown as the curved part of the surface). Also all three are zero at $t=0$ (the base of the surface shown) because a zero duration diffusion does not admit any other winding numbers than $m=n=0$, making the $Q$'s zero. However there is a spatial boundary condition either side of the half-lines $M$ and $N$, or rather a jump discontinuity condition, which suffices. The sufficiency is derived in the appendix; the jumps in the $Q$'s need evaluating.

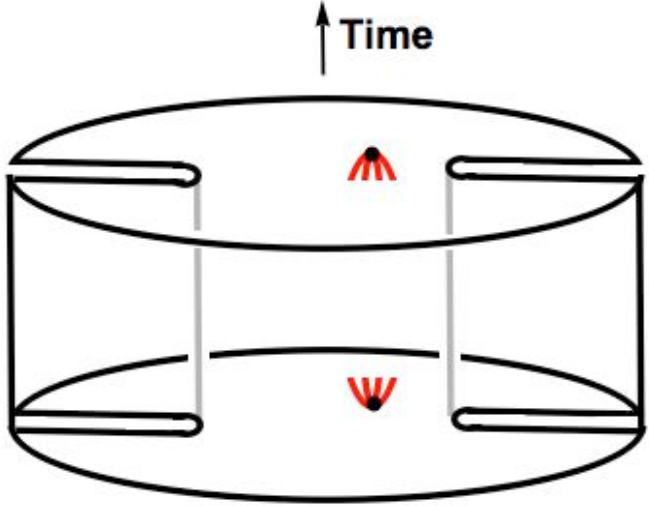

Fig 2. The surface of this doubly slotted cylinder in 2+1D spacetime is needed for evaluating the winding number correlation for Brownian loops (not shown). Actually the cylinder has infinitely large radius, and infinitely narrow slots surrounding, and excluding, the half-lines (or half-planes in spacetime) *M* and *N* (see the appendix). The 'height' of the cylinder is the duration *t* of the loop walk, and the gap between one slot and the other is the separation *L* of the marked points. Features that enter later in the calculation are symbolized by the parabolic brushes: a pair of aligned points, one a source and the other a sink of ordinary diffusion, lying respectively on the initial and final planes. Between the source and sink, the (exact) result (17) for the winding number correlation involves one 'scatter' off each of *M* and *N* as indicated in the 'top view' diagram, fig3.

For *t*>0 the quantity $Q_{MN}$ has a jump across *M* and across *N*, because $P_{mn}$ does. The quantity $Q_M$ has a jump across *M* only, because the summation over all values of *n* means that *N* plays no role – it can be considered absent. Likewise $Q_N$ has a jump across *N* only. The case of $Q_M$ will be described in detail first (and $Q_N$ is equivalent), then that of $Q_{MN}$ which, though a little more involved, follows a similar procedure.

The jump of $Q_M$ across *M* (that is, across $x'<-L/2$, $y=0$ at any chosen time $t'>0$) in the direction of increasing *y* is given by the undisturbed free space field at $(x',0,t')$ namely $P(x'-X,0-Y,t')$. The justification is straightforward using the continuity $P_{mn}(X,Y,0 \to x',0_-,t') = P_{m+1\,n}(X,Y,0 \to x',0_+,t')$. The jump in $Q_M$ is:

$$\sum_{m,n} mP_{mn}(X,Y,0 \to x',0_+,t') - mP_{mn}(X,Y,0 \to x',0_-,t') =$$

$$\sum_{m,n} mP_{mn}(X,Y,0 \to x',0_+,t') - mP_{m+1\,n}(X,Y,0 \to x',0_+,t') =$$

$$\sum_{m,n} mP_{mn}(X,Y,0 \to x',0_+,t') - (m-1)P_{mn}(X,Y,0 \to x',0_+,t') =$$

$$\sum_{m,n} P_{mn}(X,Y,0 \to x',0_+,t') = P(x'-X, 0-Y, t')$$

(8)

as claimed (remembering, for the last equality, that the sum has no jump (3) so the subscript + can be dropped). Also required will be the jump in the derivative $\partial Q_M / \partial y'$ evaluated at $y' = 0$ (i.e. on $M$). Following (8) with this derivative in place shows that this jump is $\partial P(X,Y,0 \to x',y',t')/\partial y'$ evaluated at $y'=0$. Incidentally this type of jump (or 'saltus') occurs in Kirchhoff diffraction optics [Kottler 1923, 1965][Baker and Copson 1987] [Hannay 1995][Hannay 2001] [Hannay 2010].

In the appendix it is shown using Green's theorem that knowledge of the jump across the slot(s), of a $Q$ function, together with the jump in its normal derivative, suffices for the determination of $Q$ everywhere. Actually the appendix is exemplified by the case $Q_{MN}$, but by deletion of the slot $N$ and the subscript $N$, it applies for $Q_M$ instead. The result at a general point $(x'',y'',t'')$ is

$$Q_M(X,Y,0 \to x'',y'',t'') = \int_0^{t''} dt' \int_M \tfrac{1}{2}\{P(x'-X, y'-Y, t'),\, P(x''-x', y''-y', t''-t')\} dx'$$

(9)

where $\{f,g\} \equiv f\,\partial g/\partial y' - g\,\partial f/\partial y'$. In particular the value of $Q_M$ on the other half-plane $N$ will be required soon for $Q_{MN}$ (this is well defined since $Q_M$ has no jump on $N$, only on $M$). Correspondingly the $Q_N$ scalar field evaluated on $M$ will be required. Notationally it will be useful to continue to associate ' with $M$ and to associate " with $N$. It is important that $y''$, and $y'$, should not be set to zero too soon since differentiation with respect to these variables is involved, for example the differentiation implicit in definition of the braces { }. For $Q_N$ one has the same as (9) with $M \to N$ and $' \leftrightarrow "$ (including in the definition of the braces).

Before proceeding it may be useful to summarize (and in the case of $Q_{MN}$, anticipate):

|  | Across $M$ | Across $N$ |
|---|---|---|
| Jump of $Q_M$ = | $P$ | 0 |
| Jump of $Q_N$ = | 0 | $P$ |
| Jump of $Q_{MN}$ = | $Q_N$ | $Q_M$ |

And the same applies with each of these four quantities replaced by their local $y$ derivatives e.g. Jump of $\partial Q_M/\partial y = \partial P/\partial y$ across $M$ etc. From these jump functions for the Q's across M and N, the Q's are obtained everywhere by integration 'in concatenation with' an appropriate new P, where this phrase refers to the braces form above. The winding number correlation is given by the spatial integral of $Q_{MN}$ with coincident initial and final positions (7).

Proceeding, the formula for the jump in $Q_{MN}$ across M is obtained by copying (8), but including a factor $n$ inside every summation

$$\sum_{m,n} mnP_{mn}(X,Y,0 \to x',0_+,t') - mnP_{mn}(X,Y,0 \to x',0_-,t') =$$

$$\ldots\ldots\ldots\ldots \quad (10)$$

$$= \sum_{m,n} nP_{mn}(X,Y,0 \to x',0_+,t') = Q_N(X,Y,0 \to x',0,t')$$

as claimed. Similarly the jump in $Q_{MN}$ across N is $Q_M(X,Y,0 \to x'',0,t'')$

From this, invoking Green's theorem again (appendix):

$$Q_{MN}(X,Y,0 \to x,y,t) =$$
$$\int_0^t dt'' \int_N \tfrac{1}{2}\{Q_M(X,Y,0 \to x'',y'',t''), P(x-x'',y-y'',t-t'')\}dx''$$
$$+ \int_0^t dt' \int_M \tfrac{1}{2}\{Q_N(X,Y,0 \to x',y',t'), P(x-x',y-y',t-t')\}dx' \quad (11)$$

Substituting for $Q_M$ and $Q_N$ gives nested braces. To write this out a temporary shorthand notation with $\mathbf{R}=(X,Y,0)$ and $\mathbf{r}=(x,y,t)$, is useful; for example $P(\mathbf{R} \to M)$ stands for $P(x'-X, y'-Y, t'-0)$ ). Also writing $\partial'_y \equiv \partial/\partial y'$ and $\partial''_y \equiv \partial/\partial y''$, and with $\Theta$ denoting the unit step function:

$$Q_{MN} = \iint_{M\ N}\int dx'\,dt'\,dx''\,dt''\,\Theta(t''-t') \times$$
$$\tfrac{1}{4}(\ [P(\mathbf{R} \to M)\partial'_y P(M \to N) - P(M \to N)\partial'_y P(\mathbf{R} \to M)]\partial''_y P(N \to \mathbf{r}) -$$
$$\partial''_y[Same\ square\ bracket]P(N \to \mathbf{r})\ ) \quad (12)$$

$+$ Same $\iint$ with $M \leftrightarrow N$, $' \leftrightarrow ''$ throughout

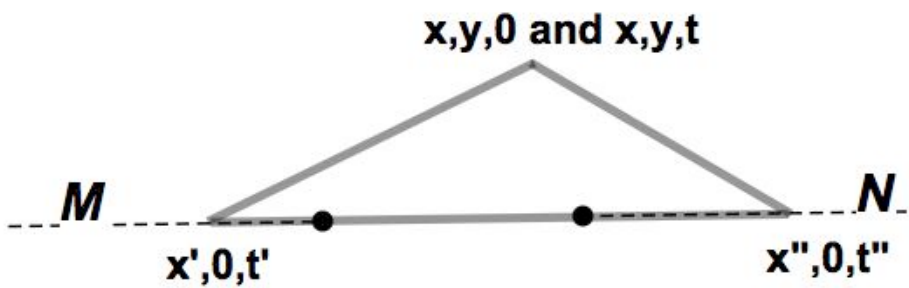

Fig 3. The two dots represent the two fixed marked points and a Brownian loop (not shown) winds randomly around them, say *n* times around one and –*m* times around the other. The winding number correlation <–*mn*> turns out to require an integration involving a product of three Gaussians whose exponents are minus half the squares of the edge lengths of a triangle divided by a duration: 0 to $t'$, $t'$ to $t''$ and $t''$ to $t$ around the triangle. (The reverse way around the triangle also contributes and doubles the answer). All six variables other than $t$ need to be integrated over, and all these integrals apart from one can be done.

Considerable simplification arises because *M* and *N* are collinear so all the first derivatives like $\partial_y P(M \to N)$ vanish, as do terms like $\partial_{y'} P(\mathbf{R} \to M)$ because *M* and *N* are separate. However the second derivative of $P(M \to N)$ does not vanish. Thus

$$Q_{MN} = \int_M \int_N dx' dt' dx'' dt'' \Theta(t''-t') \times$$

$$\tfrac{1}{4}\left(-P(M \to N)\partial'_y P(\mathbf{R} \to M)\partial''_y P(N \to \mathbf{r}) - P(\mathbf{R} \to M)\partial'_y \partial''_y P(M \to N)P(N \to \mathbf{r})\right)$$

$$+ \text{ Same with } M \leftrightarrow N, \ ' \leftrightarrow ''$$

(13)

Inserting the coordinates, and taking the initial point (*X*,*Y*) coincident with the final one (*x*,*y*),

$$Q_{MN}(x,y,0 \to x,y,t) = \int_0^t dt'' \int_0^{t''} dt' \int_{-\infty}^{-L/2} \int_{L/2}^{\infty} \frac{dx'dx''}{2^3 \pi^3 t'(t''-t')(t-t'')} \times$$

$$\tfrac{1}{4}\left(-\left(-\tfrac{\partial}{\partial y}\right)Exp\left[-\tfrac{(x-x')^2+y^2}{2t'}\right]Exp\left[-\tfrac{(x'-x'')^2}{2(t''-t')}\right]\left(-\tfrac{\partial}{\partial y}\right)Exp\left[-\tfrac{(x-x'')^2+y^2}{2(t-t'')}\right]\right.$$

$$\left. - Exp\left[-\tfrac{(x-x')^2+y^2}{2t'}\right]\tfrac{1}{(t''-t')}Exp\left[-\tfrac{(x'-x'')^2}{2(t''-t')}\right]Exp\left[-\tfrac{(x-x'')^2+y^2}{2(t-t'')}\right]\right)$$

$$+ \text{ Same with } ' \leftrightarrow ''$$

(14)

The factor $1/(t''-t')$ comes from $\partial/\partial y' \partial/\partial y'' Exp\left[-(y'-y'')^2/2(t''-t')\right]$ at $y' = y'' = 0$. Also, the single differentiations with respect to $y'$ and $y''$ have been transferred to $y$. Finally $Q_{MN}$ needs to be integrated over the *x*,*y* plane; actually this integration is done first, and then the $x'$ and $x''$ ones. It is clear that, once integrated, the piece *Same with* $' \leftrightarrow ''$ equals the rest, so it can be replaced by a factor 2. The formula is easier to read with substitutions $a = t'$, $b = t''-t'$, $c = t-t''$ (with, therefore, $a+b+c = t$). From (7) the desired correlation is now obtained. Because of the possible wider significance of the integral itself in (7), the 'normalization' factor 12, which depended on the ensemble (being chosen as the natural tractable one), is moved to the left:

$$\langle -mn \rangle / 12 = -\int Q_{MN}(x,y,0 \to x,y,t)\,dx\,dy =$$
$$2\int_0^t dt'' \int_0^{t''} dt' \frac{a+b+c}{b^2(a+c)^2} \int_{-\infty}^{-L/2} \int_{L/2}^{\infty} \frac{1}{4} Exp\left[-\frac{(a+b+c)(x'-x'')^2}{2b(a+c)}\right] \frac{dx'\,dx''}{4\pi^2} \quad (15)$$

$$= 2\int_0^t dt'' \int_0^{t''} dt' \frac{1}{4b(a+c)\pi^2} \times$$
$$\frac{1}{4}\left(Exp\left[-\frac{a+b+c}{2b(a+c)}L^2\right] - \sqrt{\pi}L\sqrt{\frac{a+b+c}{2b(a+c)}} Erfc\left[L\sqrt{\frac{a+b+c}{2b(a+c)}}\right]\right) \quad (16)$$

Using $a+b+c = t$ and $\beta = b/t$ the final result is that the winding correlation $\langle -mn \rangle$ is given by

$$\langle -mn \rangle / 12 = -\int Q_{MN}(x,y,0 \to x,y,t)\,dx\,dy$$
$$= \frac{1}{8\pi^2}\int_0^1 d\beta \frac{1}{\beta} \times \quad (17)$$
$$\left(Exp\left[-\frac{1}{2\beta(1-\beta)}\frac{L^2}{t}\right] - \sqrt{\pi}\sqrt{\frac{1}{2\beta(1-\beta)}\frac{L^2}{t}} Erfc\left[\sqrt{\frac{1}{2\beta(1-\beta)}\frac{L^2}{t}}\right]\right)$$

As would be expected, it is a monotonically decreasing function of the single variable $L^2/t$. Its integral over all separations $L$, with or without radial weighting with $L$, is simple. Indeed, weighted with the general (non-negative) integer power $L^j$, one has

$$-\int_0^\infty L^j \left[\int Q_{MN}(x,y,0 \to x,y,t)\,dx\,dy\right] dL = \frac{(j+1)\,\Gamma(\tfrac{1}{2}j+\tfrac{1}{2})^3}{32\pi^2(j+2)\,\Gamma(j+2)}(\sqrt{2t})^{j+1} \quad (18)$$

For close together marked points, specifically for $L^2 \ll t$, the bracket function in the integrand of (17) is near unity except near $\beta = 0$ and 1. Because of the $1/\beta$ outside the bracket, it is the $\beta = 0$ behaviour that matters and $(1-\beta)$ can be approximated by unity. Indeed if all the three occurrences of $(1-\beta)$ in the bracket are replaced by 1, the integral can be evaluated analytically giving the approximate form of (17) (with $\Gamma$ as the incomplete Gamma function, and $\gamma = 0.5772156...$ as Euler's Gamma constant):

$$\langle -mn \rangle / 12 \approx \frac{1}{8\pi^2}\left[\Gamma(0, L^2/2t) - 2Exp[-L^2/2t] + 2\sqrt{\pi}\sqrt{L^2/2t}\ Erfc[\sqrt{L^2/2t}]\right]$$

$$\approx \frac{1}{8\pi^2}\left[-Log\left(\frac{L^2}{2t}\right) - 2 - \gamma\right]$$

(19)

For far apart marked points, $L^2 \gg t$, the values of $\beta$ near $\beta = \frac{1}{2}$ (the saddle point) dominate the integral. Then the asymptotic expression for the bracket in the integrand of (17) is $Exp[-s^2]/2s^2$ where $s$ is the argument of the Erfc. One can set $\beta = \frac{1}{2}$ everywhere except in the expansion of $s^2$ about $\beta = \frac{1}{2}$ in the exponent, giving a Gaussian. For $L$ large, then:

$$\langle -mn \rangle / 12 \approx \frac{1}{8\pi^2}\frac{t}{2L^2}Exp\left[-\frac{2L^2}{t}\right]\int_{-\infty}^{\infty}Exp\left[-\frac{8L^2(\beta - \frac{1}{2})^2}{t}\right]d\beta$$

$$= \frac{1}{8\pi^2}\frac{\sqrt{\pi}}{2}\left(\frac{t}{2L^2}\right)^{3/2}Exp\left[-\frac{2L^2}{t}\right]$$

(20)

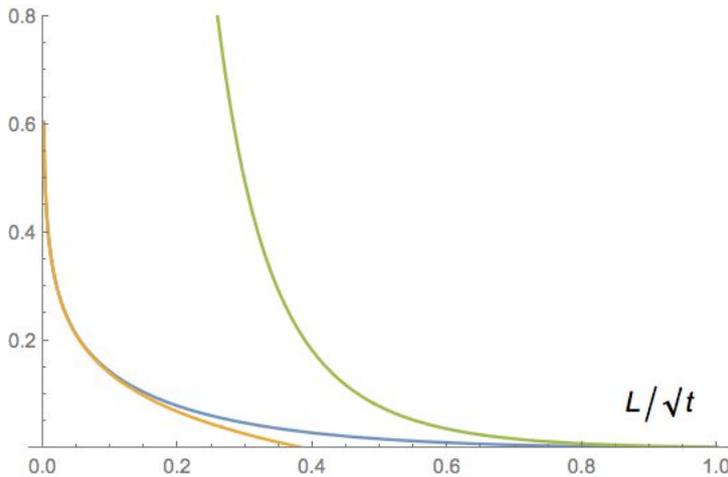

Fig 4. The graph of winding number correlation <-mn>(÷3) against the scaled separation $L/\sqrt{t}$ of the marked points (middle curve), with its approximations for small and large separations.

*Concluding remarks*
•The full topology (homotopy) of the Brownian loop in the twice punctured plane is, of course, only partially described by the winding numbers *m* and *n*. The order in which the windings happen can matter [Giraud, Thain and Hannay 2004]. For example a loop in the shape of the 'Pochhammer' complex contour (used for the Hypergeometric function) has *m*=*n*=0, which, however, cannot be 'detached' from the marked points. (The shape is that got by making a long thin figure of eight by twisting a rubber band (a loop), and bending the whole 8 into a circle so the ends just overlap. Let one marked point lie in the overlap region, and the other at the centre of the circle). Returning to the homotopy, this full topological information is not always needed and *m* and *n* do suffice for another natural description of the topology: homology.  An example is given the Aharonov-Bohm remark below.

•The normalization factor 12 in (7) arose from using the same ensemble for the two-marked-points case as was used for the single marked point case, namely Brownian loops of duration *t* with non-detachment from that point (i.e. excluding those with *n*=0).  One might aspire to use a larger ensemble, for example excluding only those loops having *m*=*n*=0.  Or larger still, excluding only those loops which are topologically fully detachable from both marked points (implying, but not implied by, *m*=*n*=0 as described in the preceding remark). Unfortunately neither of these alternative aspirations seems at all easy, (though the *m*=*n*=0 one might in principle be tractable in view of the next but one remark).  As hinted earlier, the normalization may be redundant for some purposes, as, for example, in the next remark.

• As a Feynman path integral over all paths (here written $\oiint$  ), the expression for the numerator of (7) whose evaluation is the result (17) is

$$\int dX\, dY \oiint \delta(x(0)-X)\delta(y(0)-Y)\delta(x(t)-X)\delta(y(t)-Y)\exp\left(-\tfrac{1}{2}\int_0^t (\dot{x}(\tau)^2+\dot{y}(\tau)^2)d\tau\right)\times$$

$$\int_0^t \frac{\left(x(\tau)-\tfrac{L}{2}\right)\dot{y}(\tau)-y(\tau)\dot{x}(\tau)}{\left(x(\tau)-\tfrac{L}{2}\right)^2+y(\tau)^2}\frac{d\tau}{2\pi}\ \int_0^t \frac{\left(x(\tau)+\tfrac{L}{2}\right)\dot{y}(\tau)-y(\tau)\dot{x}(\tau)}{\left(x(\tau)+\tfrac{L}{2}\right)^2+y(\tau)^2}\frac{d\tau}{2\pi}\ \frac{d^\infty x(\tau)d^\infty y(\tau)}{(2\pi t/\infty)^\infty}$$

(21)

where the last two integration terms are simply the winding numbers *n* and (−*m*).  Their product could be rewritten (perversely) in terms of auxiliary variables $\mu$ and $\nu$ as $-(\partial^2/\partial\mu\partial\nu)[\exp(i\nu\int for\ n)\exp(i\mu\int for\ (-m)]$ evaluated at $\mu=\nu=0$.  The purpose of this rewriting is that, omitting the $-(\partial^2/\partial\mu\partial\nu)$ and the $\mu=\nu=0$, the expression is that for the quantum partition function of a particle of mass $\mathcal{M}$ at temperature *T* in the presence of two Aharonov-Bohm flux points with dimensionless strengths $\mu$ and $\nu$.  The time *t* is replaced by $\hbar^2/\mathcal{M}kT$ where *k* is Boltzmann's constant (this combination has dimensions length² as it should). Thus the result (17) can be interpreted as the mixed second derivative at zero fluxes of this partition function.

•It should be mentioned here that Bogomolny [Bogomolny 2016], generalizing a method of Myers [Myers 1965] in optics, has been able to represent the quantum two-flux Aharonov-Bohm field exactly in terms of a Painlevé equation by a long calculation.  Presumably, by using properties of this equation, the partition function of the previous remark, and hence the result (17), could in principle be obtained in an alternative manner, albeit rather involved.  Indeed, though the individual probability densities $P_{mn}$ were not needed for the winding number correlation, they too might be accessible, in principle, in terms of a Painlevé equation.

•A three dimensional Brownian loop's winding number correlation around two straight parallel lines is given by the same formula (17) if the unit of time is chosen so that  <Δ$x^2$>=<Δ$y^2$>=<Δ$z^2$>=Δ$t$.  This follows because that 3D Brownian motion projected onto a plane perpendicular to the straight lines is the 2D one analysed.  (There would be cancelling factors in the ratio (7), infinity from the straight line length, and $1/\sqrt{2\pi t}$ since 2D closure does not imply 3D closure).

•It is tempting (but seems impractically difficult) to attempt Monte-Carlo confirmation of the winding number correlation by generating Brownian loops numerically.  A random Brownian loop of duration *t* can be simply constructed following Lévy [Ito and McKean 1965 page 19], by iterated polygon refinement as follows.  Successively, for 'generations' *j*=1,2…, polygons of $2^j$ vertices are defined, supplying the points of the Brownian motion at times spaced by $t/2^j$. For example the 4-gon (*j*=2) has its vertices defining the four points on the walk at 0, ¼$t$, ½$t$, ¾$t$.  To refine this to an 8-gon (the 4-gon with additional intermediate vertices at ⅛$t$, ⅜$t$, ⅝$t$, ⅞ $t$) each of the four steps separately is split into two sub-steps, the intermediate daughter point being located at a Gaussian random vector displacement from the midpoint of its parent points.  The width of the Gaussian for generating the $2^{j-1}$ additional points for the $2^j$-gon is given by <Δ$x^2$>=<Δ$y^2$>= $t/2^j$.  This is how the Brownian loop of fig 1 was generated. However, high winding numbers in random finite polygons are notoriously rare therefore giving unreliable statistics.  This defect only very slowly improves with polygon refinement iteration number *j*.  The difficulty should probably be expected given the fractal nature of Brownian motion.  My attempts to economize strategically on the refinement process proved inadequate.

•Since the work [Giraud, Thain, and Hannay 2004] has been cited several times in the present work, it seems an appropriate opportunity here to correct a minor but confusing error in fig 1 of that paper.  The top two pictures in the middle column should be grouped together in the brace, and the third should be alone, not braced with the second.

*Acknowledgement*
This calculation was prompted by a query from Michael Wilkinson who was interested in the winding number correlation <*mn*> for an application.  It had not previously occurred to me to try to calculate this (based on my earlier constructs in optics), nor that it had a rather privileged status.  It is a pleasure to thank him.

*Appendix: Boundary integral for concatenated diffusions*

Let scalar fields *A* and *B* be solutions respectively of the diffusion equation and its time reverse in 2+1D spacetime (*x*,*y*,*t*), that is, $\frac{1}{2}A_{xx} + \frac{1}{2}A_{yy} = A_t$ and $\frac{1}{2}B_{xx} + \frac{1}{2}B_{yy} = -B_t$. Then in terms of differential forms, there is a 'closed' two-form, that is, one whose exterior derivative ($\tilde{d}$) vanishes (zero div) so that its integral over any (empty, simply connected) closed spacetime surface *S* is zero:

$$\int_S \tfrac{1}{2}(AB_x - BA_x)(\tilde{d}y \wedge \tilde{d}t) + \tfrac{1}{2}(AB_y - BA_y)(\tilde{d}t \wedge \tilde{d}x) + AB(\tilde{d}x \wedge \tilde{d}y) = 0 \qquad (A1)$$

The exterior derivative of the two-form in question (the integrand) is

$$\left(\tfrac{1}{2}AB_{xx} - \tfrac{1}{2}BA_{xx} + \tfrac{1}{2}AB_{yy} - \tfrac{1}{2}BA_{yy} + AB_t + BA_t\right)(\tilde{d}x \wedge \tilde{d}y \wedge \tilde{d}t) = 0 \times (\tilde{d}x \wedge \tilde{d}y \wedge \tilde{d}t) \qquad (A2)$$

using the diffusion equations just mentioned – it constitutes Green's theorem for the diffusion equation, or similarly for the quantum Schrödinger equation or paraxial optics [Ott 1971]. The time reverse diffusion equation that *B* obeys (instead of the ordinary diffusion equation that *A* obeys) arises because *A* and *B* are to represent concatenated propagations: *A* is followed by *B*, their total duration being *t*. As the duration of *A* increases, for fixed *t*, the duration of *B* decreases.

The surface *S* suitable for the present geometry is that of fig 2; a cylinder whose axis is in the time direction, with two narrow slots to exclude *M* and *N* (which are half-planes in spacetime) so that *S* is empty of obstruction. In the circumstances relevant for the present problem the scalar field *A* represents $Q_{MN}$, and *B* represents the ordinary Gaussian diffusive sink field *P* corresponding to the upper brush in fig 2. Since $Q_{MN}$ is zero on the base and curved walls (at infinity) of *S*, only the slot faces and the 'lid' of *S* contribute to the surface integral (A1). These contributions come respectively from the middle and last term in the integral. In the last term *B* is a $\delta$-function, thereby selecting the value of $Q_{MN}$ sought.

A crucial simplifying feature of the present geometry with its infinitesimally separated flat slot faces either side of *M* and *N*, is that to evaluate the surface integral it suffices to know merely the jump of *A*, and the jump of its *y* derivative, across *M* and *N*, rather than the individual values on either side. This is because the normal directions at corresponding points either side of *M* (and of *N*) are opposite, so the combined pair in the integration picks out the *difference* of the prefactor expression $(AB_y - BA_y)$ on either side. Since the functions *B* and $B_y$ have no jump across *M* or *N*, but the function *A* does have a jump (across one or both), the combined pair gives $(A_+ - A_-)B_y - B(A_{y+} - A_{y-})$ (where the ± subscripts refer to the two sides. The jumps of *A*, and of its *y* derivative, that is, of $Q_{MN}$, are described in the main text.